\documentclass{article}

\usepackage{arxiv}

\usepackage[T1]{fontenc}    
\usepackage{url}            
\usepackage[ toc, page, title, titletoc ]{appendix}
\usepackage[ruled,noline]{algorithm2e}
\usepackage{amsmath}
\usepackage{amsfonts}
\usepackage{booktabs}
\usepackage[colorinlistoftodos]{todonotes}
\usepackage{hyperref}
\usepackage{lineno}
\usepackage{mathrsfs}
\usepackage{float}
\usepackage{url}            

\usepackage{etoolbox}
\providetoggle{fullBackground}

\usepackage{parskip}  

\DeclareFontFamily{U}{wncy}{}
\DeclareFontShape{U}{wncy}{m}{n}{<->wncyr10}{}
\DeclareSymbolFont{mcy}{U}{wncy}{m}{n}
\DeclareMathSymbol{\comb}{\mathord}{mcy}{"58} 

\usepackage{accents}
\newlength{\dhatheight}

\title{Estimating Sensitivity Maps for X-Nuclei\\ Magnetic Resonance Spectroscopic Imaging}

\author{
  Nicholas Dwork\thanks{www.nicholasdwork.com, nicholas.dwork@cuanschutz.edu} \\
  Departments of Biomedical Informatics \\
  University of Colorado Anschutz
    \And
  Jeremy W. Gordon \\
  Department of Radiology and Biomedical Imaging \\
  University of California San Francisco
    \And
  Shuyu Tang \\
  Department of Radiology and Biomedical Imaging \\
  University of California San Francisco
    \And
  Peder E. Z. Larson \\
  Department of Radiology and Biomedical Imaging \\
  University of California San Francisco
}

\begin{document}
\maketitle

\begin{abstract}
  \textbf{Purpose: }  The purpose of this research is to estimate sensitivity maps when imaging X-nuclei that may not have a significant presence throughout the field of view.

  \textbf{Methods: }  We propose to estimate the coil's sensitivities by solving a least-squares problem where each row corresponds to an individual estimate of the sensitivity for a given voxel.  Multiple estimates come from the multiple bins of the spectrum with spectroscopy, multiple times with dynamic imaging, or multiple frequencies when utilizing spectral excitation.

  \textbf{Results: }  The method presented in this manuscript, called the L2 Optimal method, is compared to the commonly used RefPeak method which uses the spectral bin with the largest magnitude to estimate the sensitivity maps.  The L2 Optimal method yields more accurate sensitivity maps when imaging a numerical phantom and is shown to yield a higher signal-to-noise ratio when imaging the brain, pancreas, and heart with hyperpolarized pyruvate as the contrast agent.

  \textbf{Conclusion: }  The L2 optimal method is able to better estimate the sensitivity by extracting more information from the measurements.
\end{abstract}


\section{Introduction}
\label{sec:intro}

When performing magnetic resonance spectroscopy of hydrogen in clinical applications, the abundant and distributed water present in the body can be used to determine the sensitivity maps throughout the field of view.  When imaging X-nuclei (atoms other than hydrogen) with custom coils, though, the sensitivity to water is negligible (since the coils are tuned to suppress its signal).  In these situations, the sensitivity maps must be determined using the X-nuclei themselves.  Example applications include imaging of phosphorus \cite{argov2000insights,hattingen2009phosphorus} or hyperpolarized $^{13}C$ \cite{brindle2011tumor,wang2019hyperpolarized,kurhanewicz2019hyperpolarized}.  For these applications, there are two categories of image reconstructions \cite{zhu2019coil}:  1)  reconstruct the image of each coil separately without knowledge of the sensitivities and then combine the images, and 2) estimate the sensitivity maps and use these maps to reconstruct a final image.

Methods of the first category, that do not require sensitivity maps, include root-sum-of-squares \cite{roemer1990nmr}, noise decorrelated combination (nd-comb) \cite{martini2010noise}, and whitened singular value decomposition (WSVD) reconstruction \cite{rodgers2010receive}.  These methods preclude generalized least squares \cite{an2013combination} or model-based \cite{fessler2010model} reconstructions.  Model-based reconstruction methods may yield improved image quality when the sensitivity maps are estimated accurately.

When sensitivity maps are desired to implement an algorithm from the second class, a common approach is to estimate them with the RefPeak method \cite{hall2014methodology,panda2012phosphorus}.  The RefPeak method uses the frequency with highest magnitude to estimate the sensitivity maps.
This method does not consider the values of other frequencies, and so it neglects to take full advantage of all of the information present in the received spectrum.  Consider the following: the magnitude of the spectrum of each voxel were the same.  Then, by averaging rather than using the maximum, (assuming independent identically distributed additive noise) one would reduce the variance in the estimate by a factor equal to the number of bins in the spectrum.  This is the best possible situation; the true benefit is a weighted average of the relative magnitudes in the spectrum.  But, as we show in this manuscript, taking advantage of this additional information is still beneficial\footnote{An early version of this work was first presented at the annual symposium of the ISMRM in 2022.}.

\section{Methods}
\label{sec:methods}

This section discusses the theory of both the RefPeak method and the novel L2 optimal method.  For both methods, estimating sensitivity values separates across voxels.
For these discussions, let $v^{(c)}\in\mathbb{C}^P$ denote the spectrum emitted by the isochromat of a given voxel and for the $c^{\text{th}}$ coil, where there are $P$ elements in the spectrum and there are $C$ coils.  Let $\rho\in\mathbb{C}^C$ denote a vector of coil sensitivity values such that $\rho_1\,v^{(1)} + \rho_2\,v^{(2)} + \cdots \rho_C\,v^{(C)}$ is proportional to the spectrum of the emitted signal (with the same constant of proportionality for every voxel).

\subsection{The RefPeak Method}

The RefPeak method identifies the maximum magnitude of the received spectrum: $A_c = \text{max}\left( v^{(c)} \right)$.  The estimated sensitivity maps are taken to be the conjugate of these maximum values, normalized to ensure consisent scaling across voxels:
\begin{equation}
  \rho_c = \frac{ A_c^\ast }{ \sqrt{ \sum_{c=1}^C |A_c|^2 } },
  \label{eq:refPeak}
\end{equation}
where $\cdot^\ast$ represents the complex conjugate.  Equation \ref{eq:refPeak} is the RefPeak method.

\subsection{The L2 Optimal Method}

Here, we present the novel L2 Optimal method.
We assume that the coil sensitivity values are those that satisfy the following for each frequency bin $\lambda$ \cite{bydder2002combination,larsson2003snr}:
\begin{equation}
  \rho_c = \frac{ v^{(c)}_\lambda }{ \left( \sum_{\gamma=1}^C \left| v^{(\gamma)}_\lambda \right|^2 \right)^{1/2} }
  \hspace{1em} \Leftrightarrow \hspace{1em}
  v^{(c)}_\lambda = \rho_c \, \left( \sum_{\gamma=1}^C \left| v^{(\gamma)}_\lambda \right|^2 \right)^{1/2}
  \label{eq:sensitivityAssumption}
\end{equation}
Since $\rho_c$ is the same for all frequency bins, the equations of \eqref{eq:sensitivityAssumption} form a system of linear system of equations with a single scalar unknown as follows:
\begin{equation}
  \begin{bmatrix}
    \left( \sum_{\gamma=1}^C \left( v^{(\gamma)}_1 \right)^2 \right)^{1/2} \\
    \left( \sum_{\gamma=1}^C \left( v^{(\gamma)}_2 \right)^2 \right)^{1/2} \\
    \vdots \\
    \left( \sum_{\gamma=1}^C \left( v^{(\gamma)}_N \right)^2 \right)^{1/2}
  \end{bmatrix} \rho_c = \begin{bmatrix}
    v^{(c)}_1 \\ v^{(c)}_2 \\ \vdots \\ v^{(c)}_N
  \end{bmatrix},
  \label{eq:linSystem}
\end{equation}
where there are $N$ bins in the spectrum.  Let $\boldsymbol{a}$ and $\boldsymbol{v}$ be defined such that \eqref{eq:linSystem} is $\boldsymbol{a} \rho_c = \boldsymbol{v}$.  Then $\rho_c$ can be estimated by minimizing $\|\boldsymbol{a} \rho_c - \boldsymbol{v}\|_2$.  This problem can be solved analytically with the pseudo-inverse: $\rho_c = ( \boldsymbol{a} \cdot \boldsymbol{v} ) / \|\boldsymbol{a}\|_2^2 $.  (Here, $\cdot$ represents the dot product.)  Estimating the sensitivity map values using \eqref{eq:linSystem} is the L2 Optimal method.

Notably, this technique is appropriate whenever there are multiple estimates with multiple coils when imaging X-nuclei.  As presented, $\lambda$ is an index into the spectrum; but it can also index time with dynamic imaging, it can index substrate when utilizing spectral-spatial excitation pulses, and it can index into combinations of these estimates.  We present results for all of these cases in section \ref{sec:results}.

\section{Experiments}
\label{sec:experiments}

We evaluated the novel L2 Optimal method and compared it to the RefPeak method with a numerical phantom, and data of a heart, a brain, and a pancreas.  Human data was acquired of volunteers using hyperpolarized $^{13}C$ with a 3 Tesla clinical MRI scanner.

\begin{figure}[ht]
  \centering
  \includegraphics[width=0.6\textwidth]{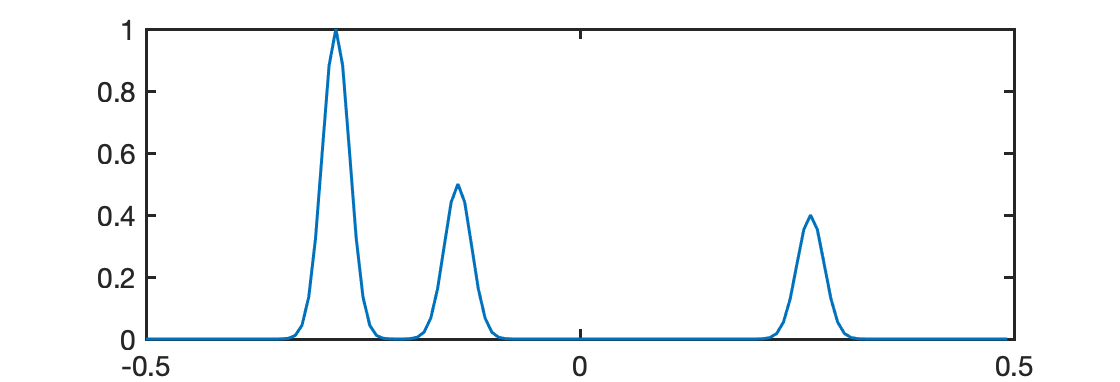}
  \caption{Spectrum for numerical phantom.}
  \label{fig:phantomSpec}
\end{figure}

The numerical phantom was generated using the Shepp-Logan phantom \cite{shepp1974fourier} imaged with a simulated 8-coil birdcage sensor, a Cartesian sampling trajectory with two spatial dimensions of phase encodes and readout in the spectral dimension.  All pixels exhibited the same spectrum, that shown in Fig. \ref{fig:phantomSpec}, and the intensity was apodized by the intensity of the phantom.  Eight rectangular coils were used in the simulation; they were evenly spaced around the brain with a distance of $0.5$ meters between opposite coils.  The Biot-Savart law was used to simulate the sensitivity maps for each coil \cite{esin2017mri}.  Coil coupling was simulated by constructing the sensitivity matrix $\mathcal{S}$ (the $i^\text{th}$ column of $S$ is a column extension of the $i^\text{th}$ sensitivity map) and projecting it onto the closest matrix (in a Frobenius sense) of rank $5$.  The first row of Fig. \ref{fig:phantomMaps} shows the simulated sensitivity maps.

\textit{The MR data of the head, pancreas, and heart were gathered with Institutional Review Board (IRB) approval, Health Insurance Portability and Accountability Act (HIPAA) compliance, and patient informed assent/consent.  These images were de-identified and no protected health information was provided to the engineering team.}

For the data of the head and the pancreas, custom coils were used to image hyperpolarized $[1-^{13}C]$ pyruvate after injection \cite{wang2019hyperpolarized,kurhanewicz2019hyperpolarized}.  For the head and pancreas, the Echo-Planar Spectroscopic Imaging (EPSI) \cite{gordon2017development} trajectory was used:  with each excitation, a single line of $k$-space was acquired with an echo-planar trajectory.  The heart was imaged with a $5$ slice spatial-spectral acquisition where pyruvate, lactate, and bicarbonate were each acquired individually with a spiral acquisition.  For both the pancreas and the heart, images were acquired approximately every $2$ seconds to observe the temporal dynamics of the bolus injection.

\section{Results}
\label{sec:results}

\begin{figure}[ht]
  \centering
  \includegraphics[width=0.8\textwidth]{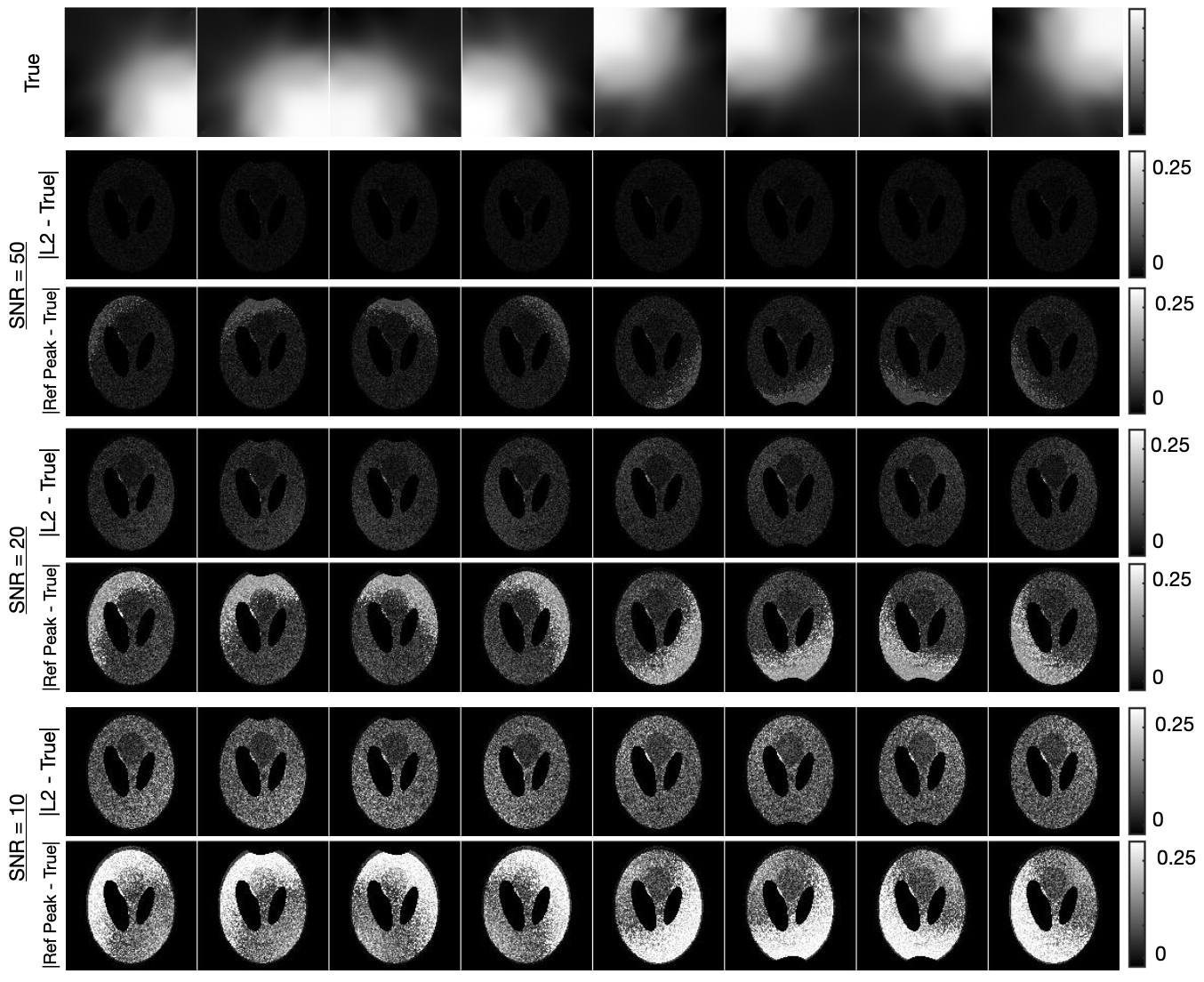}
  \caption{The first row shows the true sensitivity maps.  All other rows show the difference between truth, the estimates attained with the L2 Optimal method, and the Ref Peak method for various signal to noise ratios.  Differences are shown on a scale from $0$ to $0.25$.  In all cases, the estimates made by the L2 Optimal method are more accurate.}
  \label{fig:phantomMaps}
\end{figure}

Figure \ref{fig:phantomMaps} shows the results of estimating sensitivity maps with the numerical phantom for different signal-to-noise (SNR) ratios.  The first row shows the true sensitivity maps, and the subsequent rows show the difference between the estimates and truth.  Points in the phantom that were vacant of signal were assigned a sensitivity map value of $0$.  For all SNR, the L2 Optimal method has significantly reduced error than the RefPeak method.  This is quantified in table \ref{table:mse}, where the mean square error of each set of estimates was calculated.  The mean square error of the L2 Optimal method is approximately $5-10$ times lower than that of the RefPeak method.

\begin{table*}[ht]
  \centering
  \caption {Mean Squared Error of Sensitivity Map Estimates}
  \begin{tabular}{lcccc}
    \toprule
    SNR &  Inf & 50 & 20 & 10 \\
    \midrule
    Ref Peak  & < $10^{-16}$ & $1.2\cdot 10^{-3}$ & $1.1\cdot 10^{-2}$ & $4.0\cdot 10^{-2}$  \\
    L2 Optimal & < $10^{-16}$ & $1.5\cdot 10^{-4}$ & $1.5\cdot 10^{-3}$ & $9.1\cdot 10^{-3}$  \\
    \bottomrule
  \end{tabular}
  \label{table:mse}
\end{table*}

\begin{figure}[ht]
  \centering
  \includegraphics[width=0.9\textwidth]{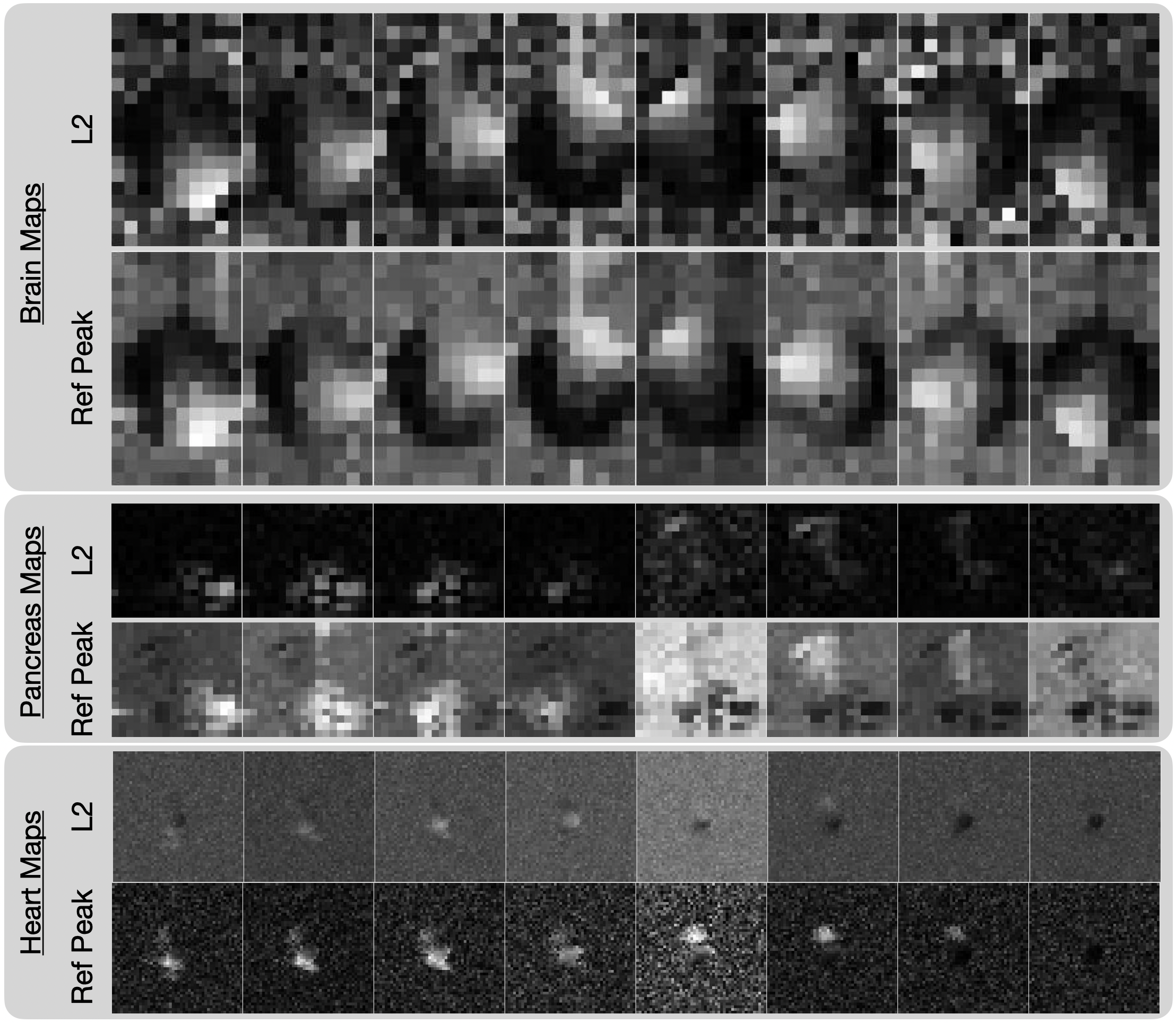}
  \caption{Sensitivity maps and reconstructions of the brain, pancreas, and heart using the RefPeak method and the L2 Optimal methods.  Note that the sensitivity maps of the L2 Optimal method and the RefPeak method are shown on the same intensity scale.}
  \label{fig:sMaps}
\end{figure}

Figure \ref{fig:sMaps} shows the sensitivity maps estimated for the brain (top) and the pancreas (bottom).  In regions where there is signal, the estimate of the sensitivity maps in the brain are not significantly different between the L2 Optimal method and the RefPeak method.  This is because the intensity of pyruvate is much higher than any bin in the rest of the spectrum; thus, there is little additional information to gain from the remaining portions of the spectrum.  However, when solving the least-squares problem based on \eqref{eq:linSystem} with the pseudo-inverse, the result is the smallest vector that minimizes the objective.  Thus, the regions of the image where there was no sample (outside of the brain) have smaller values with the L2 Optimal method than with the RefPeak method.

For the pancreas, an EPSI trajectory was used with a temporal dynamic acquisition, where images were acquired approximately every $2$ seconds.  In this case, for the L2 Optimal method, the $\lambda$ of \eqref{eq:linSystem} is an index into both frequency and time.  The spectrum has the same property as the brain (the pyruvate signal intensity dominates the spectrum), but the multiple images provide an opportunity for an improved estimate.  The result, as seen in Fig. \ref{fig:sMaps}, are sensitivity map estimates with less noise.

\begin{figure}[ht]
  \centering
  \includegraphics[width=0.9\textwidth]{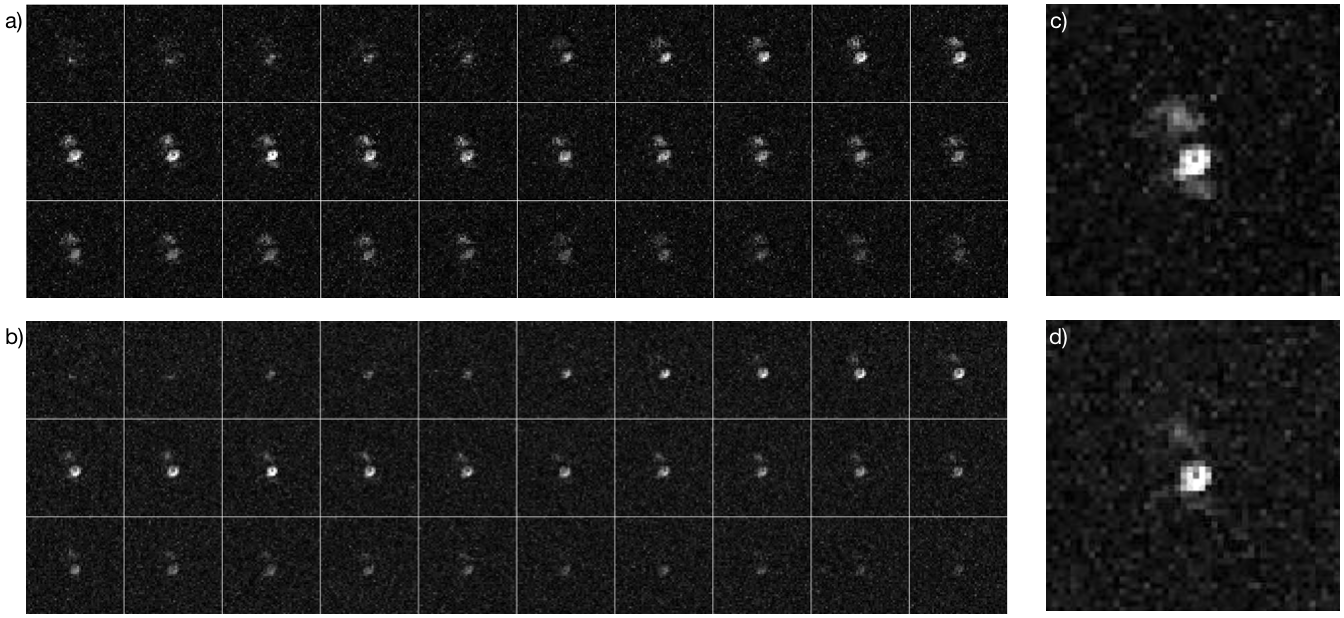}
  \caption{Images of pyruvate in the third slice of heart with L2 Optimal (top) and RefPeak (bottom) methods.  Subfigures (a) and (b) show $30$ images separated by approximately $2$ seconds.  Subfigures (c) and (d) show fourteenth images enlarged.}
  \label{fig:heartReconsPyr}
\end{figure}

\begin{figure}[ht]
  \centering
  \includegraphics[width=0.9\textwidth]{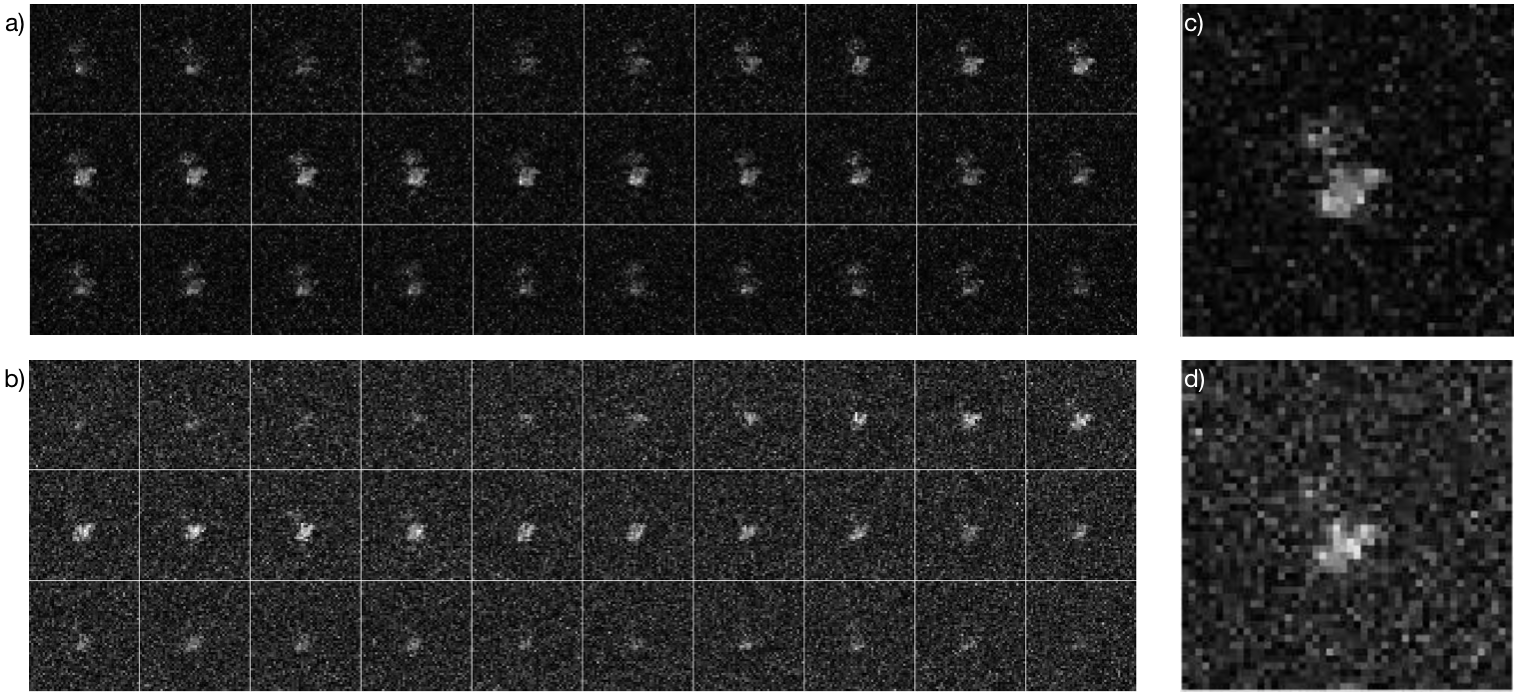}
  \caption{Images of bicarbonate in the third slice of heart with L2 Optimal (top) and RefPeak (bottom) methods.  Subfigures (a) and (b) show $30$ images separated by approximately $2$ seconds.  Subfigures (c) and (d) show tenth images enlarged.}
  \label{fig:heartReconsBic}
\end{figure}

Images from all coils were combined into a single image using the method of Roemer \cite{roemer1990nmr}.  Figures \ref{fig:heartReconsPyr} and \ref{fig:heartReconsBic} show reconstructions of the pyruvate and bicarbonate in the center slice of the heart, respectively.  (Reconstructions of lactate are not shown in this manuscript for conciseness, but data from lactate was included in the analysis.)  In this case, the $\lambda$ of \eqref{eq:linSystem} indexed into metabolite and time.  The SNR of the reconstructions for bicarbonate are significantly improved.  For both pyruvate and bicarbonate, though, the reconstructions are significantly different suggesting that the L2 Optimal method offers an improved understanding of the spatial metabolic activity.

\begin{figure}[ht]
  \centering
  \includegraphics[width=0.6\textwidth]{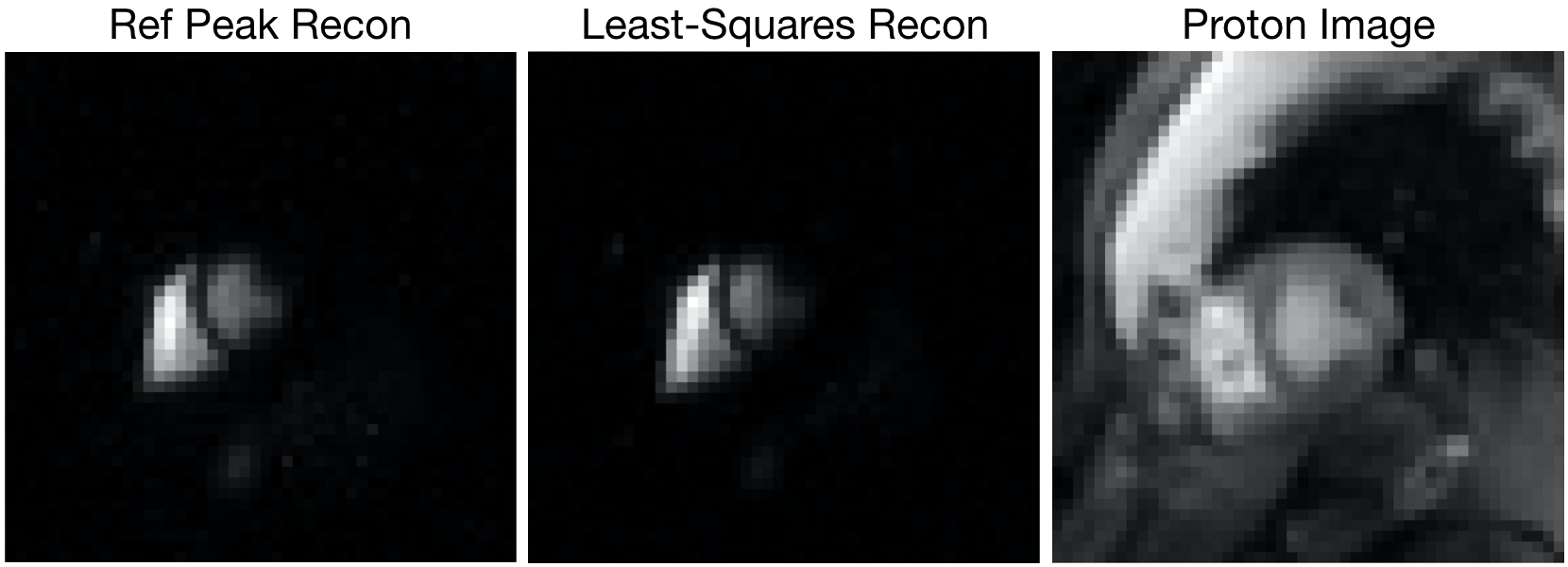}
  \caption{Images of cardiac pyruvate after an injection of [1- $^{13}C$] with the (left) RefPeak and (center) least-squares method.  The right image shows a corresponding anatomical slices of the heart.}
  \label{fig:pyrCardiac}
\end{figure}

Figure \ref{fig:pyrCardiac} shows a slice of the heart where individual coil images have been combined using the method of Roemer \cite{roemer1990nmr} with sensitivity maps generated from the RefPeak and L2 Optimal method; these images are placed beside a corresponding proton image.  Note that there is a significant difference between these images, and that the least-squares image better localizes the signal of pyruvate in the myocardium.

\section{Discussion and Conclusions}
\label{sec:conclusion}

Minimizing the L2 norm is optimal when experiencing additive independent, identically distributed, Gaussian noise.
In this manuscript, we compared the L2 Optimal method of estimating sensitivity maps (presented in this manuscript) to the RefPeak method for imaging x-nuclei and have shown that the L2 Optimal method more accurately estimates the sensitivities, reduces the estimate of the sensitivity when the signal of the sample is negligible, and significantly changes the reconstruction of the image.  The improvement in the reconstruction is more pronounced when the signal is low (e.g. when imaging bicarbonate in a hyperpolarized [$1-C^{13}$] experiment), but still changes the reconstruction when the signal is high.  These improvements are a result of exploiting the information in multiple measurements of the sensitivity for each voxel.

\section*{Acknowledgments}
ND has received post-doctoral training funding from the American Heart Association (grant number 20POST35200152).
ND has received funding support from the Quantitative Biosciences Institute at UCSF (no grant number).
PL has received funding support from the the National Institute of Health (grant numbers NIH R01 HL136965 and R33 HL161816).
PL has received funding support from the American Cancer Society (grant number 131715-RSG-18-005-01-CCE).
JG has received funding from the National Institute of Health / National Institute of Biomedical Imaging and Bioengineering (grant number U01EB026412).

PL and JG have also received research support from GE Healthcare, which may be considered a conflict of interest.



\begin{thebibliography}{10}

\bibitem{argov2000insights}
Zohar Argov, Mervi L{\"o}fberg, and Douglas~L Arnold.
\newblock Insights into muscle diseases gained by phosphorus magnetic resonance
  spectroscopy.
\newblock {\em Muscle \& nerve}, 23(9):1316--1334, 2000.

\bibitem{hattingen2009phosphorus}
Elke Hattingen, J{\"o}rg Magerkurth, Ulrich Pilatus, Anne Mozer, Carola
  Seifried, Helmuth Steinmetz, Friedhelm Zanella, and R{\"u}diger Hilker.
\newblock Phosphorus and proton magnetic resonance spectroscopy demonstrates
  mitochondrial dysfunction in early and advanced parkinson's disease.
\newblock {\em Brain}, 132(12):3285--3297, 2009.

\bibitem{brindle2011tumor}
Kevin~M Brindle, Sarah~E Bohndiek, Ferdia~A Gallagher, and Mikko~I Kettunen.
\newblock Tumor imaging using hyperpolarized 13c magnetic resonance
  spectroscopy.
\newblock {\em Magnetic resonance in medicine}, 66(2):505--519, 2011.

\bibitem{wang2019hyperpolarized}
Zhen~J Wang, Michael~A Ohliger, Peder~EZ Larson, Jeremy~W Gordon, Robert~A Bok,
  James Slater, Javier~E Villanueva-Meyer, Christopher~P Hess, John
  Kurhanewicz, and Daniel~B Vigneron.
\newblock Hyperpolarized {13C} {MRI}: state of the art and future directions.
\newblock {\em Radiology}, 291(2):273--284, 2019.

\bibitem{kurhanewicz2019hyperpolarized}
John Kurhanewicz, Daniel~B Vigneron, Jan~Henrik Ardenkjaer-Larsen, James~A
  Bankson, Kevin Brindle, Charles~H Cunningham, Ferdia~A Gallagher, Kayvan~R
  Keshari, Andreas Kjaer, Christoffer Laustsen, et~al.
\newblock Hyperpolarized {13C} {MRI}: path to clinical translation in oncology.
\newblock {\em Neoplasia}, 21(1):1--16, 2019.

\bibitem{zhu2019coil}
Zihan Zhu, Xucheng Zhu, Michael~A Ohliger, Shuyu Tang, Peng Cao, Lucas
  Carvajal, Adam~W Autry, Yan Li, John Kurhanewicz, Susan Chang, et~al.
\newblock Coil combination methods for multi-channel hyperpolarized {13C}
  imaging data from human studies.
\newblock {\em Journal of Magnetic Resonance}, 301:73--79, 2019.

\bibitem{roemer1990nmr}
Peter~B Roemer, William~A Edelstein, Cecil~E Hayes, Steven~P Souza, and
  Otward~M Mueller.
\newblock The {NMR} phased array.
\newblock {\em Magnetic Resonance in Medicine}, 16(2):192--225, 1990.

\bibitem{martini2010noise}
N~Martini, MF~Santarelli, G~Giovannetti, M~Milanesi, D~De~Marchi, V~Positano,
  and Luigi Landini.
\newblock Noise correlations and snr in phased-array {MRS}.
\newblock {\em NMR in Biomedicine: An International Journal Devoted to the
  Development and Application of Magnetic Resonance In vivo}, 23(1):66--73,
  2010.

\bibitem{rodgers2010receive}
Christopher~T Rodgers and Matthew~D Robson.
\newblock Receive array magnetic resonance spectroscopy: whitened singular
  value decomposition ({WSVD}) gives optimal bayesian solution.
\newblock {\em Magnetic Resonance in Medicine: An Official Journal of the
  International Society for Magnetic Resonance in Medicine}, 63(4):881--891,
  2010.

\bibitem{an2013combination}
Li~An, Jan Willem van~der Veen, Shizhe Li, David~M Thomasson, and Jun Shen.
\newblock Combination of multichannel single-voxel {MRS} signals using
  generalized least squares.
\newblock {\em Journal of Magnetic Resonance Imaging}, 37(6):1445--1450, 2013.

\bibitem{fessler2010model}
Jeffrey~A Fessler.
\newblock Model-based image reconstruction for {MRI}.
\newblock {\em {IEEE} signal processing magazine}, 27(4):81--89, 2010.

\bibitem{hall2014methodology}
Emma~L Hall, Mary~C Stephenson, Darren Price, and Peter~G Morris.
\newblock Methodology for improved detection of low concentration metabolites
  in mrs: optimised combination of signals from multi-element coil arrays.
\newblock {\em Neuroimage}, 86:35--42, 2014.

\bibitem{panda2012phosphorus}
Anshuman Panda, Scott Jones, Helmut Stark, Rahul~S Raghavan, Kumar
  Sandrasegaran, Navin Bansal, and Ulrike Dydak.
\newblock Phosphorus liver {MRSI} at {3T} using a novel dual-tuned
  eight-channel {31P/1H} coil.
\newblock {\em Magnetic resonance in medicine}, 68(5):1346--1356, 2012.

\bibitem{bydder2002combination}
Mark Bydder, David~J Larkman, and Joseph~V Hajnal.
\newblock Combination of signals from array coils using image-based estimation
  of coil sensitivity profiles.
\newblock {\em Magnetic Resonance in Medicine}, 47(3):539--548, 2002.

\bibitem{larsson2003snr}
Erik~G Larsson, Deniz Erdogmus, Rui Yan, Jose~C Principe, and Jeffrey~R
  Fitzsimmons.
\newblock {SNR}-optimality of sum-of-squares reconstruction for phased-array
  magnetic resonance imaging.
\newblock {\em Journal of Magnetic Resonance}, 163(1):121--123, 2003.

\bibitem{shepp1974fourier}
Lawrence~A Shepp and Benjamin~F Logan.
\newblock The fourier reconstruction of a head section.
\newblock {\em Transactions on nuclear science}, 21(3):21--43, 1974.

\bibitem{esin2017mri}
Yunus~Emre Esin and Ferda~Nur Alpaslan.
\newblock {MRI} image enhancement using {Biot}-{Savart} law at 3 {Tesla}.
\newblock {\em Turkish Journal of Electrical Engineering \& Computer Sciences},
  25(4), 2017.

\bibitem{gordon2017development}
Jeremy~W Gordon, Daniel~B Vigneron, and Peder~EZ Larson.
\newblock Development of a symmetric echo planar imaging framework for clinical
  translation of rapid dynamic hyperpolarized {13C} imaging.
\newblock {\em Magnetic resonance in medicine}, 77(2):826--832, 2017.

\end{thebibliography}

\end{document}